# Optimising RF linear Amplifier for maximum efficiency and linearity

Srđan Milić
Faculty of Electrical Engineering and Computing
University of Zagreb, HR-10000 Zagreb, Croatia
E-mail: srdan.milic@fer.hr
E-mail: igor.krois@fer.hr

*Abstract*—A method for increasing efficiency of radio frequency (RF) amplifier employing laterally diffused metal oxide semiconductor (LDMOS) transistors coupled to an RF exciter depending on the emission mode of modulated RF input signals generated by exciter, if exciter output signal is of a type where modulated RF signals do not have continuously varying envelope, biasing the LDMOS transistor in the RF amplifier with fixed quiescent drain current and fixed drain voltage supply to cause LDMOS transistors to operate in compression and if exciter output signal is of a type where modulated RF signals do have continuously varying envelope, biasing the LDMOS transistors in the RF amplifier for linear operation.

## I. Introduction

Laterally diffused metal oxide (LDMOS) transistor devices have found wide use in RF power amplifier designs requiring high output power and drain-to-source breakdown voltages in excess of 130V. Such amplifiers typically have been designed to operate at reduced output levels and consequently lower efficiency in order to maintain dependability, ruggedness and good output signal quality/linearity for signals having high peak-to-average envelope variation.

High efficiency RF power amplifiers have always been of interest to RF engineers, a large volume of literature on the subject, ranging from theoretical analyses to practical circuit designs, has been published over years [1-4]. On the theory side, most analyses utilised simplified models for the devices and circuits to make mathematics involved manageable. Due to lack of low frequency models for given device used most of the process in constructing RF linear amplifier is based on cut and try method.

This article presents design and methods involved in increasing RF amplifier efficiency depending on the emission mode of modulated RF input signal generated by exciter from the perspective of RF/electronics student.

The RF linear power amplifier evolved from a project 1kW 1.8 to 30MHz broadband LDMOS design. Average drain efficiency between 75 to 85% was achieved at required power and frequency, the circuit employs MRFX1K80H laterally diffused metal oxide semiconductor (LDMOS) transistor together with commonly used wideband transmission line (TLT) transformers for practical RF linear amplifier circuits and several tuning capacitors. Series of controlled experiments were carried out in order to test and obtain good compromise between maximum power output, maximum efficiency and minimum realistic intermodulation products (IMD3, IMD5 etc…) possible.

## II. Background and Circuit Description

The output device pressed into service is NXP semiconductors MRFX1K80H high power and ruggedness N – channel enhancement – mode lateral transistor. It has 1800W CW typical output power capability and typically operates at 65V drain bias. Its unmatched input and output design supports frequency use from 1.8 to 400 MHz, the circuit schematic and a photo of actual RF amplifier are shown in figures 1 and 2. Input matching is realised with 1:4 centre tapped ferrite loaded broadband RF transformer, output transformer utilises slightly different transformer type, instead of using conventional flux coupled transformer which has limited frequency response and needs compensation in order to properly cover 1.8 to 30 MHz frequency range, the transformer of choice is ferrite loaded transmission line (TLT) transformer offering much broader and less lossy characteristics. Because traditional transformers are not used different method for applying drain DC bias is employed in form of shunt connected bifilar drain choke, consequently, drain bias circuit becomes part of the output matching network doing extra impedance transformation.

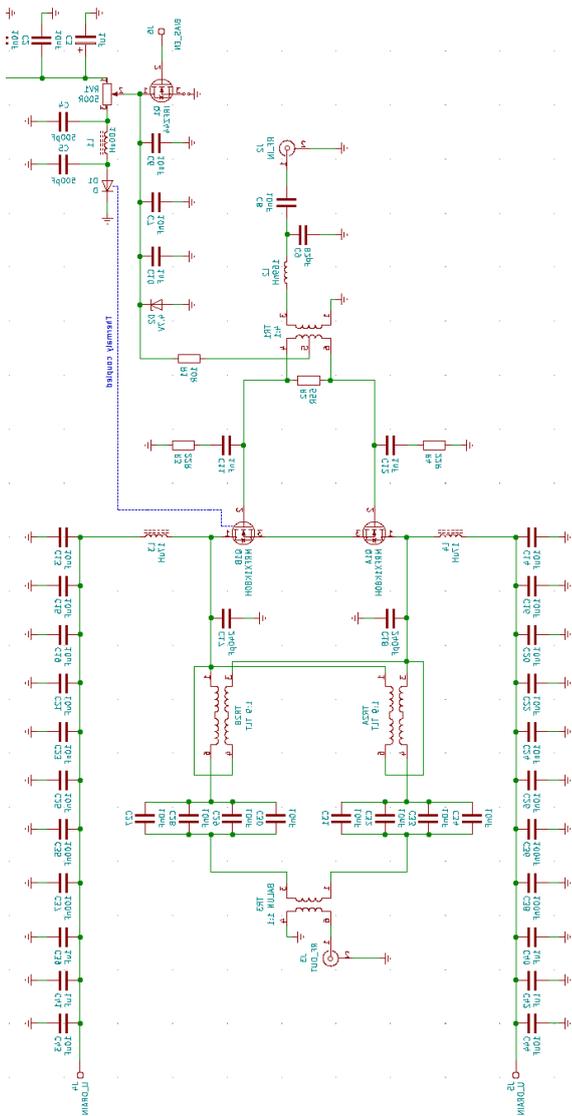

*Figure 1 RF amplifier deck schematic*

At the input transistor gates are loaded with RC snubber network in order to lower the Q and suppress parasitic oscillations in gate circuit, also author opted for no negative feedback as it could easily become positive feedback at some frequency due to lack of modelling data and previous experiments showed signs of possible instability at higher operating frequencies, input circuit is optimised for good broadband characteristics offering return loss ranging from -18dB to -22dB as the power gain of the amplifier is very high ~ 30dB so solely matching for power gain was not necessary.

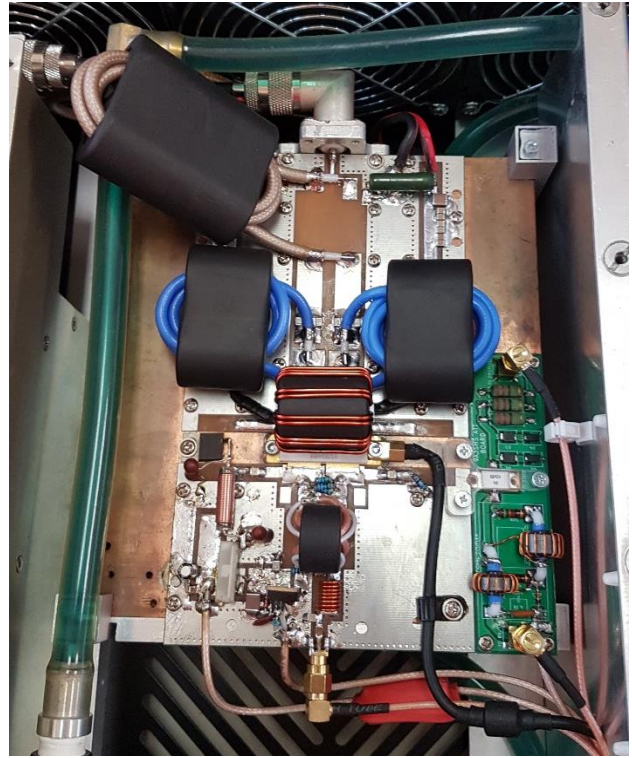

*Figure 2 photo of constructed RF amplifier deck*

Traditionally high efficiency power amplifiers are categorized into classes, such as class AB, B and C amplifiers, class F and class E amplifiers. Each class corresponds to a specific mechanism for achieving high efficiency. From class A to class AB, B and C amplifiers, the progressive improvement in efficiency is achieved with the reduced conduction angle. In this category of amplifier, the waveform of drain current is assumed to be a quasi-sine wave. The Fourier analysis of this waveform shows that the ratio of the fundamental to DC component increases as the conduction angle is reduced, resulting in increasing efficiency with the reduction of conduction angle [3].

The latter is true if we take into consideration only the conduction angle, but also efficiency of AB class amplifiers heavily depends on how hard the amplifier is driven so depending on the emission mode of modulated RF input signal generated by exciter the amplifier operation point is modified to achieve high efficiency or high linearity.

## III. MEASUREMENT RESULTS

As mentioned before this article focuses on determining exciter mode of operation thus determining if modulated signal from exciter has time varying envelope amplitude or time invariant envelope amplitude, depending on determined mode RF amplifier logic changes its bias voltages and shifts operating point accordingly, for signals comprising from constant envelope goal is to lower IDQ to 500mA and lower the drain bias so its just above amplified signal envelope.

Why we do this? By lowering drain bias and IDQ transistors are forced to operate in compression, typically 2 to 3dB in compression, output power is increased only 100 to 150W but efficiency gain is 15% so it translates to less heating enabling us to dampen the cooling requests and lowering power consumption. This is possible only for certain types of signals like FM, CW and digital. Doing this on variable envelope signals such as AM, SSB will create significant distortion and intermodulation products creating horrible splatter to neighbouring channels interfering with data transmission. For such signals RF amplifier is strictly operating in linear mode with high drain bias and IDQ of 2A which is not optimal for efficiency and power consumption but offers high linearity and low intermodulation products.

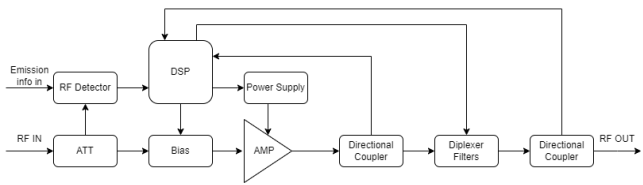

*Figure 3 RF amplifier block diagram*

| Drain bias(V) | Gain (dB) | Eff (%) | Output power(W) | Dissipation (W) |
|---|---|---|---|---|
| 58 | 32 | 60 | 1000 | 666 |
| 53 | 30 | 68 | 1000 | 470 |
| 48 | 28 | 77 | 1000 | 298 |

*Figure 4 Drain efficiency vs CW output power vs drain bias*

RF amplifiers DSP unit oversees presetting gate bias and drain bias, gate bias adjustment is realised with microprocessor, LM317 voltage regulator and 5 transistors enabling 5 step preselection of gate bias.

Drain bias supply is repurposed Huawei R4850G2 50V 55A server power supply, fine drain bias level regulation is possible via CAN interface to Huawei power supply ranging from 30V to 58V.

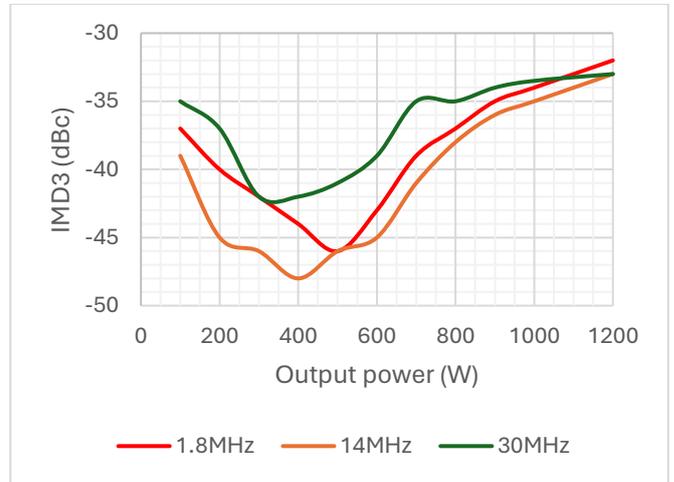

*Figure 5 Intermodulation measurement using 2-tone test*

As we can see in figure 5 IMD3 products are highly dependable on power level and at the same time from figure 6 power level is dependent on drain bias level.

So it is possible to fine tune RF amplifier gain with subtle changes in drain bias supply without significantly impacting the amplifier static operating point e.g. IDQ.

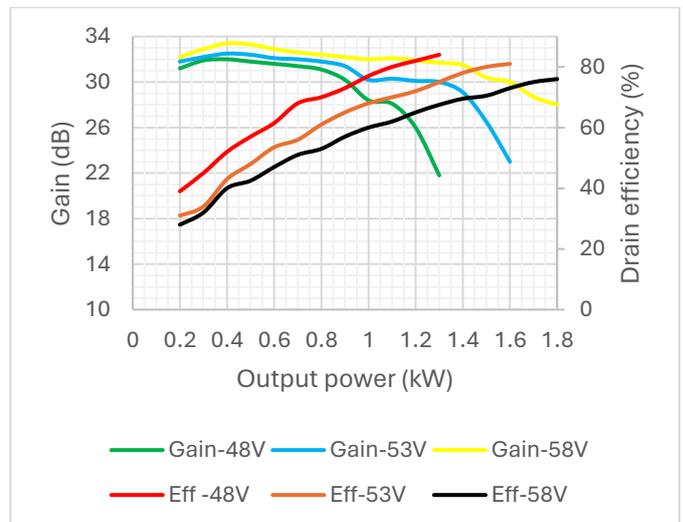

*Figure 6 Drain efficiency vs CW output power vs drain bias - detailed*

## IV. Conclusions

High power RF linear amplifiers can enhance their operating efficiency at the cost of signal linearity, if the modulated input signal from exciter has constant envelope like in FM and CW modes it is possible to improve efficiency, if the modulated input signal from exciter has varying envelope like SSB or AM signals it is not possible to improve efficiency in this matter the next step would be to use highly efficient and fast PWM modulators to modulate RF amplifier drain bias supply in order to further enhance efficiency. This method is called envelope restoration and elimination (EER) and is more complex requiring synchronisation and more processing power.

Having CAN bus programable drain bias supply, it gives us possibility to fine tune amplifier gain at certain bands as input and output networks are comprised from realistic components which introduce uneven response, as the author is using this feature to equalise gain across 160M, 80M, 60M, 40M, 30M, 20M, 17M, 15M, 12M and 10M radio amateur bands.

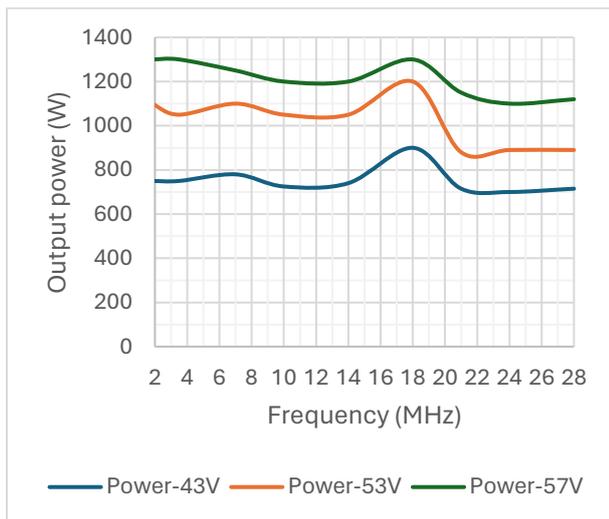

*Figure 7 RF Amplifier CW Output power vs Frequency response*

### Acknowledgemet

The author would like to thank Ranko Boca for his insight and many helpful advices during the discussions involved in this project.

The author would like to thank his mentor Igor Krois for his contribution and help during this project.